\documentclass[aps,twocolumn,pre,showpacs,color,notitlepage]{revtex4-1}
\usepackage{graphicx}
\usepackage{amsmath}
\usepackage[normalem]{ulem}
\usepackage{soul}
\usepackage[utf8]{inputenc}
\usepackage[english]{babel}
\usepackage{lipsum}
\usepackage{csquotes}

\usepackage{marginnote}
\usepackage{cooltooltips}

\usepackage{pstricks}
\newpsobject{showgrid}{psgrid}{subgriddiv=2,griddots=10,gridlabels=3mm}

\begin{document}


\title{Noise-induced standing waves in oscillatory systems with time-delayed feedback}

\author{Michael Stich}
  \thanks{Corresponding author: m.stich@aston.ac.uk} 
\author{Amit K. Chattopadhyay}
\thanks{a.k.chattopadhyay@aston.ac.uk}
  \affiliation{Non-linearity and Complexity Research Group, Systems Analytics Research Institute, School of Engineering and Applied Science, Aston University, Aston Triangle, Birmingham, B4 7ET, UK.}


\date{\today}

\begin{abstract}
  In oscillatory reaction-diffusion systems, time-delay feedback can
  lead to the instability of uniform oscillations with respect to
  formation of standing waves. Here, we investigate how the presence
  of additive, Gaussian white noise can induce the appearance of
  standing waves. Combining analytical solutions of the model with
  spatio-temporal simulations, we find that noise can promote standing
  waves in regimes where the deterministic uniform oscillatory modes
  are stabilized. As the deterministic phase boundary is approached,
  the spatio-temporal correlations become stronger, such that even
  small noise can induce standing waves in this parameter regime. With
  larger noise strengths, standing waves could be induced at finite
  distances from the (deterministic) phase boundary. The overall
  dynamics is defined through the interplay of noisy forcing with the
  inherent reaction-diffusion dynamics.
\end{abstract}

\maketitle

\section{Introduction}

Reaction-diffusion models define a paradigmatic class of systems to
study wave patterns in spatially-extended media far from thermal
equilibrium~\cite{Hoyle06}. Beyond their natural use in chemical
systems~\cite{Kapral95}, they have been applied to general
pattern-forming dynamical systems~\cite{CrossRMP93}, kinetic
roughening systems~\cite{Halpin-Healy95}, biological
systems~\cite{Murray89}, among others.

Here, we consider the case where the reaction-diffusion system has
undergone a smooth transition from a stationary state to uniform
oscillations, a scenario captured by the supercritical Hopf
bifurcation. The temporal and spatio-temporal behavior of the system
is then described by the complex Ginzburg-Landau equation
(CGLE)~\cite{CrossRMP93}. However, uniform oscillations are not the
only solution to that equation: among the most studied traveling wave
solutions are one-dimensional plane waves and two-dimensional spiral
waves. Furthermore, fascinating aspects of such dynamics concern
unstable oscillations often leading to spatio-temporal chaos, like
phase turbulence and defect
chaos~\cite{Murray89,AransonRMP02,PeregoPRL16}.  The motivation of our
work is to suppress spatio-temporal chaos in the CGLE and to replace
it with regular patterns in a stochastically forced setting.  The
underlying method with which we achieve this is time-delay feedback.

Control of chaotic states in pattern-forming systems is a wide field
of research that has already been reviewed in detail (e.g.,
in~\cite{MikhailovPR06,Scholl07}).  In the context of the
reaction-diffusion systems, the introduction of forcing terms or
global feedback terms have been shown to be efficient ways to control
turbulence. To cite just one example, chemical turbulence can be
suppressed by global time-delayed feedback~\cite{KimS01,BetaPRE03} in
the CO oxidation reaction on Pt(110). In principle, most real physical
feedbacks would need some time to influence the system. Although there
may be cases where the feedback is fast enough compared to the
intrinsic characteristic time scale and hence can be regarded as
instantaneous, in general such a feedback would act with a time delay
$\tau$. This sort of delay may appear under two heads, a spatially
dependent {\emph{local feedback}} and a spatially independent
{\emph{global feedback}}. In global feedback, a spatially-averaged
variable or a variable without space dependence is fed back to the
system dynamics.  In the context of the CGLE, global feedback with
explicit time delay was considered by Battogtokh and
Mikhailov~\cite{BattogtokhPD96} and then Beta and
Mikhailov~\cite{BetaPD04}. The latter used the Pyragas feedback
scheme, where the feedback signal is created from the difference
between the actual system state and a time-delayed
one~\cite{PyragasPLA92}. Among other features, the authors reported a
parameter regime between spatio-temporal chaos and uniform
oscillations where {\emph{standing wave patterns}} were observed.

The presence of noise changes the dynamics of nonlinear,
spatially-extended systems significantly, as noise can not only
destabilize certain patterns, but it also can enhance and induce
others, as reviewed in~\cite{GarciaOjalvo99}.  Recently, the effect of
noise on systems subjected to time delay has attracted interest, like
in the context of noise-induced oscillations~\cite{PomplunEpL05},
correlation times~\cite{BrandstetterPTRSA10}, stochastic
bifurcation~\cite{ZakharovaPRE10}, coherence
resonance~\cite{GeffertEPJB14}, stochastic
switching~\cite{DhuysPRE14}, or autonomous
learning~\cite{KaluzaPRE14}. These studies, though, primarily focus on
systems without spatial extension, whereas this article considers a
reaction-diffusion system and therefore enables us to study a
spatially-extended wave pattern under the simultaneous influence of
time delay and noise. In the context of extended systems, different
features of spatial and temporal coherence due to noise (but without
time delay) close to pattern-forming
instabilities~\cite{CarrilloEpL04}, in excitable
systems~\cite{PercPRE05}, and for coupled chaotic
oscillators~\cite{ZhouPRE02,KissC03} have been considered.  The effect
of noise on time-delay models has been studied, e.g., for a network of
excitable Hodgkin-Huxley elements~\cite{WangPLA08}.

This work builds on the foundation laid out in the seminal work by
DeDominicis and Martin~\cite{DeDominicis79,Chattopadhyay01}. Based on
a stochastically forced Burgers' dynamics, later to be followed by the
paradigmatic Kardar-Parisi-Zhang model \cite{Kardar86}, the results
highlighted the importance of stochastic forcing in second order phase
transitions~\cite{Barabasi95}. Here we take this approach one step
further, by including a finite time delay in a stochastically forced
spatio-temporal dynamics that threads together vital \enquote{missing
  links} in the causality analysis of a perturbed stochastic
dynamics. The key construct here is the segregation of the mean and
fluctuating components of a dynamical field, in line with the
DeDominicis-Martin scheme~\cite{DeDominicis79}.  The methodology has
recently been successfully used in fluid and magnetohydrodynamic
models as
well~\cite{Chattopadhyay01,Chattopadhyay13,Chattopadhyay14}. In this
approach, each vector field $\phi$ will be split into a mean component
$\phi_0$ and a stochastic random part $\delta \phi$ representing the
(often) nonlinear flow close to the boundary layer as follows: $\phi =
\phi_0 + \delta \phi$.  The component $\delta \phi$ represents the
fluctuation dominated regime away from the line of symmetry.  Such a
segregation of deterministic and stochastic components in the model
allows one to study the perturbed dynamics of $\delta \phi$ around the
mean (symmetry) variable $\phi_0$ as a set of two coupled equations,
one in $\delta \phi$ and the other in $\phi_0$.

The focal point here is the analysis of the above stochastically
forced dynamical field $\delta \phi$ in the context of time delay.  In
a series of works~\cite{BetaPD04,StichPRE07,StichPD10,StichPRE13},
time-delay feedback has been used to suppress spatio-temporal chaos in
the CGLE without stochastic terms and different aspects have been
considered, like the interplay of local vs. global feeback
terms~\cite{StichPRE07}, the stability of the uniform
solutions~\cite{StichPD10}, and the standing-wave
solution~\cite{StichPRE13}. In this work, instead of including local
feedback terms, for the sake of simplicity we use a stochastic
generalization of the model with purely global feedback, introduced in
Ref.~\cite{BetaPD04}.  In the context of our model, our interests are
in understanding the following: a) how noise modifies the transition
from a turbulent regime via standing waves to a state of uniform
oscillations, and b) whether standing waves themselves can be induced
by noise.

This paper is organized as follows: in Section~\ref{sec:model}, we
introduce the model and describe briefly the relevant deterministic
solutions, uniform oscillations and standing waves. In
Section~\ref{sec:cf}, we introduce noise terms and calculate the
spatio-temporal correlation functions.  In Section~\ref{sec:sim}, we
show numeric simulations to explore the onset of standing waves in the
presence of noise. A summary of results and future directions of
research are presented in Section~\ref{sec:disc}.

\section{The deterministic model and its main solutions}
\label{sec:model}

Reaction-diffusion systems can display various types of oscillatory
dynamics.  However, close to a supercritical Hopf bifurcation, all
such systems are described by the complex Ginzburg-Landau equation
(CGLE)~\cite{CrossRMP93},
\begin{equation}
  \frac{\partial A(x,t)}{\partial t}=(1-{\rm i}\omega )A-(1+{\rm i}\alpha )|A|^{2}A+(1+{\rm i}%
  \beta )\Delta A,  
  \label{cgl}
\end{equation}
where $A$ is the complex oscillation amplitude, $\omega$ the linear
frequency parameter, $\alpha$ the nonlinear frequency parameter,
$\beta$ the linear dispersion coefficient, and $\Delta$ stands for the
Laplacian operator. For $1+\alpha\beta<0$ (the Benjamin-Feir-Newell
criterion), uniform oscillations $A_u={\mathrm{exp}}(-{\mathrm
  i}(\omega+\alpha)t)$ are unstable and spatio-temporal chaos is
observed. In analogy with~\cite{DeDominicis79}, the $\phi_0$ there
serves the role of the spatio-temporal field variable $A(x,t)$.

The CGLE for a one-dimensional medium with global time-delayed
feedback $F$ has been introduced in Ref.~\cite{BetaPD04} and is
defined by
\begin{subequations}
\label{eq:model}
\begin{eqnarray} %
  \frac{\partial A(x,t)}{\partial t}&=&(1-{\rm i}\omega )A-(1+{\rm i}\alpha )|A|^{2}A \nonumber \\
  &+&(1+{\rm i}%
    \beta ) \frac{\partial^2 A}{\partial x^2} 
    +F,\:\:\text{and}
    \label{eq:modela}
    \end{eqnarray}
\begin{equation}
    F = \mu {\mathrm e}^{{\mathrm i}\xi} \left (
    \bar{A}(t-\tau)-\bar{A}(t) \right), 
\label{eq:modelb}
\end{equation} 
\end{subequations}
where
$\bar{A}(t)=\frac{1}{L}\int_0^L A(x,t) \, {\rm d}x$
denotes the spatial average of $A(x,t)$ over a one-dimensional medium
of length $L$.  The parameter $\mu$ describes the feedback strength
and $\xi$ characterizes a phase shift between the feedback and the
current dynamics of the system.

The solution of the {\emph{feedback-induced}} uniform oscillations is
given by $A_{UO}(t)=\rho_0 {\mathrm{exp}}(-{\mathrm i}\Omega
t)$~\cite{BetaPD04}, where the amplitude and frequency are given by
\begin{subequations}  \label{eq:uniform}%
\begin{eqnarray}
    \rho_0 &=& \sqrt{1 + \mu [\cos(\xi+\Omega\tau)-\cos\xi]}, \\
    \Omega &=& \omega + \alpha + \mu \bigg[\alpha(\cos(\xi+\Omega\tau)-\cos\xi) \nonumber \\
    &-& (\sin(\xi+\Omega\tau)-\sin\xi)\bigg]\,.
  \end{eqnarray}
\end{subequations} 
In general, no explicit analytic solution for Eqs.~(\ref{eq:uniform})
can be given. Nevertheless, the solutions can be found numerically
using root-finding algorithms. In order to understand the suppression
of spatio-temporal chaos, a linear stability analysis for uniform
oscillations was done~\cite{BetaPD04}.  At stable uniform
oscillations, control of chaos was consistently achieved. Obviously,
this depends not only on the CGLE parameters, but also on the control
parameters, in particular $\mu$ and $\tau$ (we consider a fixed $\xi$
throughout the article). In the limits where the feedback strength or
the time delay go to zero, the feedback term also goes to zero. This
makes the scheme ineffective, and spatio-temporal chaos is
recovered.

In order to analyze the stochastically forced CGLE model, the
stability boundaries of uniform oscillations in the parameter space need
to be ascertained for the deterministic model~(\ref{eq:model}). These
boundaries are given by the conditions $\lambda_1=0$ and $\partial_p
\lambda_1 \ne 0$, where $\lambda_1$ is thereal part of the dominant
eigenvalue (the others must be negative) and $p$ stands for either
$\mu$ or $\tau$. As shown in detail in~\cite{BetaPD04,StichPD10}, we
can specify the parameter sets for which the uniform periodic solution
becomes unstable with respect to \emph{standing waves} with wavelength
$2\pi/k_c$ ($k_c\ne 0$), where $k_c$ is the critical wavenumber as
given by the linear stability analysis of the uniform
oscillations~\cite{BetaPD04}. It varies between $0.7$ and $0.9$ for
the parameter set we are interested in, see Fig. 5(b)
of~\cite{BetaPD04}.

\begin{figure}[!t] 
 \centering 
\includegraphics[height=8.0cm,width=8.5cm]{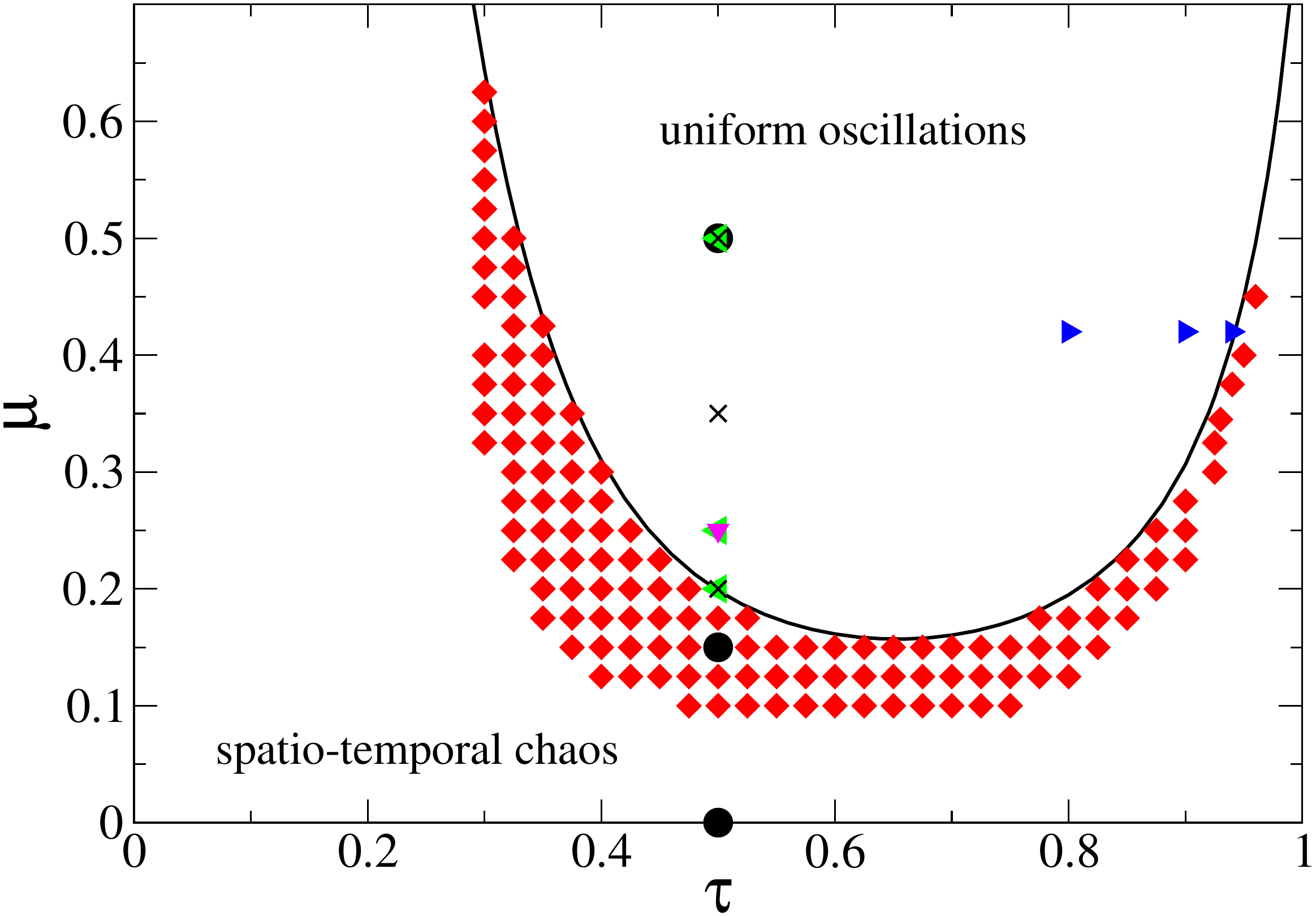}
\vspace{0.5cm}
\caption{Main solutions in the parameter space spanned by $\tau$ and
  $\mu$. The solid line defines the stability boundary of uniform
  oscillations in the deterministic system (above the curve). Below
  that curve, the diamond symbols indicate simulations displaying
  standing waves in the deterministic system (data from Fig.~8 of
  Ref.~\cite{BetaPD04} and own simulations). The circles denote the
  parameter values chosen as defined in Fig.~\ref{fig222}, while the
  left triangles indicate the parameter values as used for
  Fig.~\ref{figNEW333}(a,b); the right triangles represent the
  parameter values as used for Fig.~\ref{figNEW333}(c,d), and the down
  triangle stands for the parameter value for Fig.~\ref{fig555}. The
  crosses represent the parameter values used in
  Fig.~\ref{fig666}. Note that with the exception of
  Fig.~\ref{fig222}(b,c,d), all simulations were performed in the
  deterministically stable regime characterized by uniform
  oscillations where standing waves do not exist. The other parameters
  are: $\alpha=-1.4$, $\beta=2$, $\omega=2\pi-\alpha$, $\xi=\pi/2$.}
 \label{fig111}%
\end{figure}

In Fig.~\ref{fig111}, a part of the $\mu-\tau$ parameter
space is shown where uniform oscillations are stable (above the
solid curve), and where standing waves are found numerically
(diamonds). The other symbols indicate parameter values used in later
figures.

Simulations confirm that the onset of standing waves is smooth, and
that the standing wave is characterized by a vanishing space-dependent
part at threshold. In this model, standing wave solutions are
described by~\cite{StichPRE13}
\begin{equation}
\label{finalsol}
A_{SW}={\mathrm e}^{-{\mathrm i}\Omega_0 t} (H_0
+ 2 B_{k0} \cos(kx) {\mathrm e}^{-{\mathrm i}\gamma}),
\end{equation}
where $k$ is given by the eigenvalue problem studied
in~\cite{BetaPD04}, i.e., it corresponds either to $k_c$ (at onset of
the standing wave pattern, $\lambda_1=0$) or $k_{max}$ (away from
onset, $\lambda_1 \ne 0$), and $H_0$, $B_{k0}$, $\Omega_0$, and
$\gamma$ are given by a set of nonlinear equations given
in~\cite{StichPRE13}. This deterministic formulation will be later
used as we define the amplitude of noise-induced standing waves.

\begin{figure}[!ht] 
 \centering 
\includegraphics[height=8.0cm,width=8.5cm]{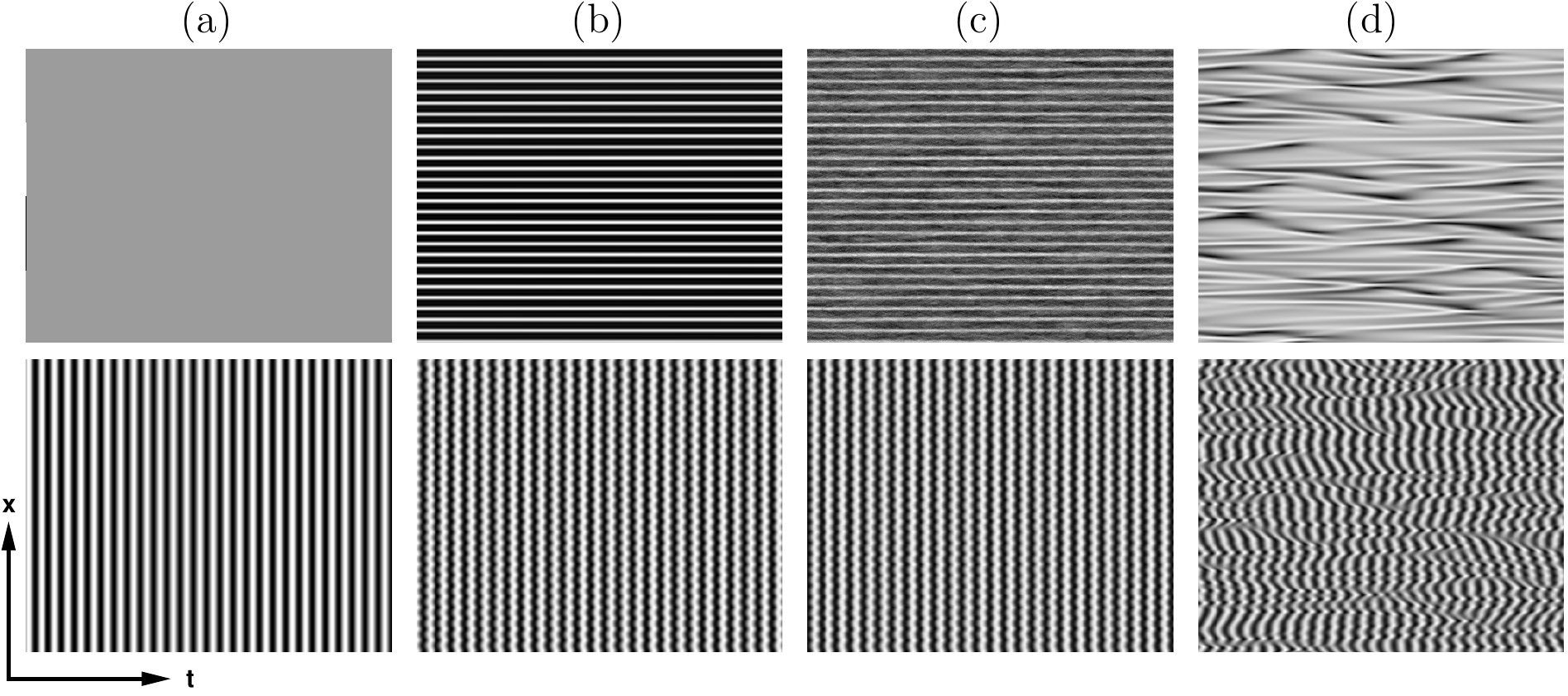}
\caption{Main spatio-temporal solutions for different feedback
  magnitudes and noise strengths: (a) uniform oscillations, (b)
  standing waves without noise, (c) standing waves with small noise,
  (d) spatio-temporal chaos.  Shown are space-time diagrams in gray
  scale for $|A|$ (top panels) and Re$A$ (bottom panels) for a time
  interval of $t=25$ in the asymptotic regime and system size $L=256$.
  The delay time is $\tau=0.5$ and the values of $\mu$ are $\mu=0.50$
  (a), $\mu=0.15$ (b), $\mu=0.15$ (c), $\mu=0$ (d). The noise
  magnitude is $D=0.05$ in (c) and zero otherwise.  Black (white)
  denotes low (high) values of the respective quantity (rescaled for
  each simulation). For $|A|$, these values are
  $(|A|_{min},|A|_{max})=(0.94,1.13)$ (b),
  $(|A|_{min},|A|_{max})=(0.9,1.15)$ (c),
  $(|A|_{min},|A|_{max})=(0.15,1.2)$ (d). For (a), the amplitude is
  constant $|A|=1.085$. The other parameters are as in
  Fig.~\ref{fig111}.}
 \label{fig222}%
\end{figure}

Spatio-temporal simulations are performed for a one-dimensional system
with size $L=256$ and spatial resolution $\Delta x=0.32$. For time
integration, we use an explicit Euler scheme with $\Delta t =
0.002$. The Laplacian operator is discretized using a next-neighbor
representation, as discussed for the deterministic model used
in~\cite{StichPRE13} (and references therein).  We apply periodic
boundary conditions and the initial conditions consist of developed
spatio-temporal chaos as present in the absence of feedback. Usually,
the system settles to an asymptotic state before $t = 200$, while we
let it evolve until $t = 500$. Then, we start the simulations that are
shown in Figs.~\ref{fig111},~\ref{fig222},~\ref{fig555},~\ref{fig666}.

In Fig.~\ref{fig222}, we give an overview of the most relevant
patterns, as observed in the simulations of the model defined in
Eq.~(\ref{eq:model}).  The upper panels show space-time diagrams of
$|A|$, the lower panels representing the solutions for the real part
of the amplitude. The latter illustrates the oscillations, while the
former reveal the amplitude of the oscillations and whether they have
a space dependence. According to the Benjamin-Feir-Newell criterion,
the Ginzburg-Landau parameters $\alpha$ and $\beta$ are chosen to
fulfill $1+\alpha\beta<0$, i.e., in the absence of feedback, the
system converges to the regime of spatio-temporal chaos. This is shown
in space-time diagrams for $|A|$ and Re$A$ (Fig.~\ref{fig222}(d)),
where Re$A$ denotes the real part of $A$. But in the presence of
strong feedback ($\mu=0.5$), the feedback induces uniform oscillations
(Fig.~\ref{fig222}(a)). For an appropriate choice of the delay time
$\tau$, between the chaotic region and the region of uniform
oscillations, standing waves are observed. As $\mu$ decreases (for
this $\tau$, at $\mu_c=0.19848$), small-amplitude standing waves set
in (Fig.~\ref{fig222}(b)). These standing waves are spatial
modulations of the underlying uniform oscillations. For comparison
with the stochastic model discussed below (Section~\ref{sec:sim}), we
show in Fig.~\ref{fig222}(c) the impact of small noise to the standing
waves (otherwise same parameters as in Fig.~\ref{fig222}(b)). If the
noise is small enough, the observed pattern is stable and clearly
recognizable, in spite of inevitable small fluctuations.

\section{The stochastic model and its correlation functions}
\label{sec:cf}

While previous works~\cite{BetaPD04,StichPRE13} gave us an
understanding of standing waves in the deterministic system, the
dynamics of these waves in the presence of noise and in particular
their onset are unknown.  In order to tackle this question, we analyze
the stochastic Langevin model, starting from
equations~(\ref{eq:model}). This can be accomplished by studying the
impact that the spatio-temporal noise $N(x,t)$ has on the system, in
particular when we approach the instability of uniform oscillations
with respect to perturbations with $k\ne0$. Model~(\ref{eq:model})
therefore becomes
\begin{subequations}
\label{eq:modelnoise}%
\begin{eqnarray}  
    \frac{\partial A}{\partial t} &=& (1-{\rm i}\omega )A-(1+{\rm i}\alpha )|A|^{2}A\nonumber \\
    &+&(1+{\rm i}%
    \beta ) \frac{\partial^2 A}{\partial x^2} 
    +F+N(x,t),\\
     \langle N(x,t) N(x',t') \rangle  &=& 2D\delta(x-x')\delta(t-t'),
\end{eqnarray} 
\end{subequations}
where $N(x,t)$ stands for a Gaussian, white noise with magnitude $D$,
and where $F$ is given in Eq.~(\ref{eq:modelb}). In order to
calculate the correlation functions, we resort to a Fourier series
expansion of $N(x,t)$ as follows
\begin{equation}
N(x,t) = \int\tilde{N}_{k,\tilde{\omega}}\,{\mathrm e}^{{\mathrm i}(kx-\tilde{\omega} t)}{\mathrm d}k\,{\mathrm d}\tilde{\omega}.
\label{fouriertransforms}
\end{equation}
For $A$, we use the ansatz
\begin{equation}
A(x,t)=\rho_0\exp(-{\mathrm i}\Omega t)+A_{+}\exp({\mathrm i}kx)+A_{-}\exp(-{\mathrm i}kx),
\end{equation}
where $A_{\pm}$ represent the amplitudes of the linearly independent
solutions $\exp(\pm {\mathrm i}kx)$, phenomenologically representing
oppositely directed waves from left to right or from right to
left. The wave vector $k$ is determined from linear stability
analysis, details of which are available in~\cite{BetaPD04}. Our
interest is in the spatio-temporal autocorrelations of the field $A$
that will allow us to compare and establish the contributions from
stochasticity driven perturbations against the results obtained in the
previous non-noisy cases~\cite{BetaPD04,StichPRE07,StichPD10}. The
necessary quantities to calculate in this connection are respectively
the autocorrelation function $C_0=\langle A(x,t)*A^{*}(x,t) \rangle$,
the spatial correlation function $C_r=\langle
{[A^{*}(x+r,t)-A(x,t)]}^2 \rangle=2(C_0- \langle
A^{*}(x+r,t)*A(x,t)\rangle )$ and the temporal correlation function
$C_{t'}=\langle {[A^{*}(x,t+t')-A(x,t)]}^2 \rangle=2(C_0-\langle
A^{*}(x,t+t')*A(x,t)\rangle )$. The brackets denote ensemble
averages. Straightforward algebra then leads us to the following
results:
\begin{subequations}
\begin{eqnarray}
C_0 &=& 2D \bigg[ {|\rho_0|}^2 xt + \frac{4}{k(\lambda_1^2 + \lambda_2^2)}\text{Re}(\rho_0 {A_+}^{*(0)}+\rho_0 {A_-}^{*(0)}) \nonumber \\ 
&\times& \sin\left(\frac{kx}{2}\right) \bigg[{\mathrm e}^{\lambda_1 t}\left(\lambda_1 \cos(\lambda_2 t+\frac{kx}{2})+\lambda_2  \sin(\lambda_2 t+\frac{kx}{2})\right) \nonumber \\ 
&-& \lambda_1\cos(\frac{kx}{2})-\lambda_2 \sin(\frac{kx}{2})\big] \bigg] 
\label{autocorr_eqn} 
\end{eqnarray}
\begin{eqnarray}
C_r &=& \frac{8D}{k({\lambda_1}^2 + {\lambda_2}^2)}\bigg[ \text{Re}(\rho_0 {A_+}^{*(0)}+\rho_0 {A_-}^{*(0)})e^{\lambda_1 t}
\big[ \sin(\frac{kx}{2}) \nonumber \\
&\times&(\lambda_1 \cos(\lambda_2 t+\frac{kx}{2})+ \lambda_2 \sin(\lambda_2 t+\frac{kx}{2})) 
- \sin(\frac{k}{2}(x+r)) \nonumber \\
&\times& (\lambda_1 \cos(\lambda_2 t+\frac{k}{2}(x+r))+ \lambda_2 \sin(\lambda_2 t+\frac{k}{2}(x+r))) \big] \nonumber \\
&\times&\lambda_1[\cos(\frac{kx}{2})-\cos(\frac{k}{2}(x+r))] \nonumber \\
&+& \lambda_2[\sin(\frac{kx}{2})-\sin(\frac{k}{2}(x+r))] \bigg]
  \label{spatialcorr_eqn}
 \end{eqnarray}
 \begin{eqnarray}
C_{t'} &=& 2\bigg[ C_0 - 2D {\mathrm e}^{{\mathrm i} \Omega t'}\big[{|\rho_0|}^2 xt + {\rho_0}^{*} A_{+}^{(0)} \left(\frac{1-{\mathrm e}^{{\mathrm i}kx}}{{\mathrm i}k\lambda}\right)\left(1-{\mathrm e}^{\lambda t}\right) \nonumber \\
&-& {\rho_0}^{*} A_{-}^{(0)} \left(\frac{1-{\mathrm e}^{-{\mathrm i}kx}}{{\mathrm i}k\lambda^{*}}\right) \left(1-{\mathrm e}^{\lambda^{*} t}\right)  \nonumber \\
&-& {\rho_0} {A_{+}}^{*(0)} \left(\frac{1-{\mathrm e}^{-{\mathrm i}kx}}{{\mathrm i}k\lambda^{*}}\right) \left(1-{\mathrm e}^{\lambda^{*} t}\right) \nonumber \\
&+& {\rho_0} {A_{-}}^{*(0)} {\mathrm e}^{\lambda t'} \left(\frac{1-{\mathrm e}^{{\mathrm i}kx}}{{\mathrm i}k\lambda}\right) \left(1-{\mathrm e}^{\lambda t}\right)\big] \bigg],
\label{tempcorr_eqn}
\end{eqnarray}
\end{subequations}
where $\lambda=\lambda_1 + {\mathrm i} \lambda_2 $ and
$\lambda^{*}=\lambda_1 -{\mathrm i}\lambda_2 $, $\lambda_{1,2}$ being
the solutions of the quadratic equation $\lambda^2
-2(1-k^2{\rho_0}^2)\lambda+[1+\omega^2+2\beta \omega k^2 + 4\alpha
  \omega {\rho_0}^2 + (1+\beta^2) k^4 + 4\alpha \beta {\rho_0}^2 k^2 +
  3(1+\alpha^2) {\rho_0}^4]=0$ (see Ref.~\cite{BetaPD04}) and
$\rho_0$ and $\Omega$ are given by Eqs.~(\ref{eq:uniform}). Note that
$\lambda_1$ denotes here the real parts of the eigenvalues of the
linear stability analysis of uniform oscillations, as explained above.

\begin{figure}[h] 
 \centering 
\includegraphics[height=8.0cm,width=8.5cm]{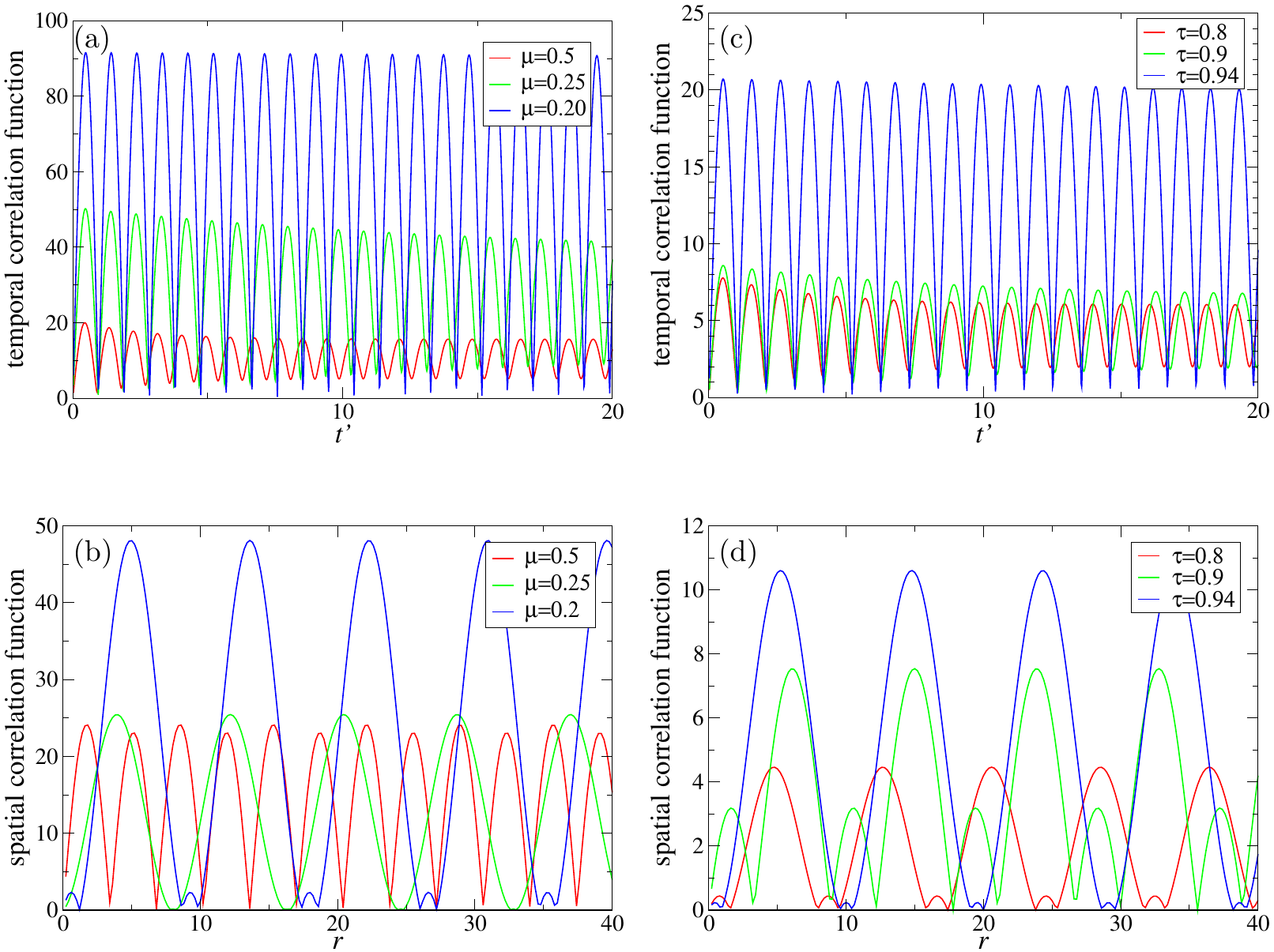}
\caption{(a,b) Amplitudes of temporal (a) and spatial (b) correlation
  functions for $\tau=0.5$ for three different values of $\mu$ (see
  legend) and $D=1$. We observe that the closer $\mu$ to the critical
  $\mu_c=0.19848$, the larger the magnitude of the correlation
  functions. For illustration, we rescale (multiply) the temporal
  correlation functions with 1000 ($\mu=0.5$) and 50 ($\mu=0.25$) and
  the spatial correlation functions with 100 ($\mu=0.5$) and 50
  ($\mu=0.25$). (c,d) Amplitudes of temporal (c) and spatial (c)
  correlation functions for $\mu=0.42$ for three different values of
  $\tau$ (see legend) and $D=1$. We observe that the closer $\tau$ to
  the critical $\tau_c=0.94244$, the larger the magnitude of the
  correlation functions. For illustration, we rescale (multiply) the
  correlation functions with 10 ($\tau=0.8$) and 5 ($\tau=0.9$).  All
  other parameters are as in Fig.~\ref{fig111}.}
 \label{figNEW333}
\end{figure}

In this context, spatial and temporal correlation functions are of
particular interest.  In Fig.~\ref{figNEW333}(a,b), we observe the
amplitude of the spatial ($C_r$) and temporal ($C_{t'}$) correlation
functions for a fixed $\tau$ as we approach the instability of uniform
oscillations and the simultaneous onset of standing waves (at
$\mu_c=0.19848$). The influence of the noise can be expected to be
more prominent as we approach the instability and hence the magnitude
of the correlation functions should increase towards the
instability. This is exactly what is observed in
Fig.~\ref{figNEW333}(a,b) for three different parameter values.  To
show different evaluations of the correlation functions in the same
figure, we have rescaled the correlation functions (see figure
captions).  Since the solution describes temporal oscillations, they
are also present in the temporal correlation functions
(Fig.~\ref{figNEW333}(a)). We see that away from the instability
($\mu=0.5$), the temporal correlation function approaches a constant
envelope value after approximately 20 time units.  On the other hand,
the spatial correlation function (Fig.~\ref{figNEW333}(b)) does not
show a decaying property as the temporal one, and the periodicity
corresponds to the $k$ value resulting from the linear stability
analysis~\cite{BetaPD04}.

In Fig.~\ref{figNEW333}(c,d), we show the correlation functions (as
functions of $r$ and $t'$ respectively) for three values of $\tau$
while keeping $\mu=0.42$ constant. Qualitatively, we observe a similar
behavior as in Fig.~\ref{figNEW333}(a,b). As the delay time $\tau$
increases towards its critical value, the amplitude of the correlation
functions also increases.

\begin{figure}[h] 
 \centering 
\includegraphics[height=5.0cm,width=8.5cm]{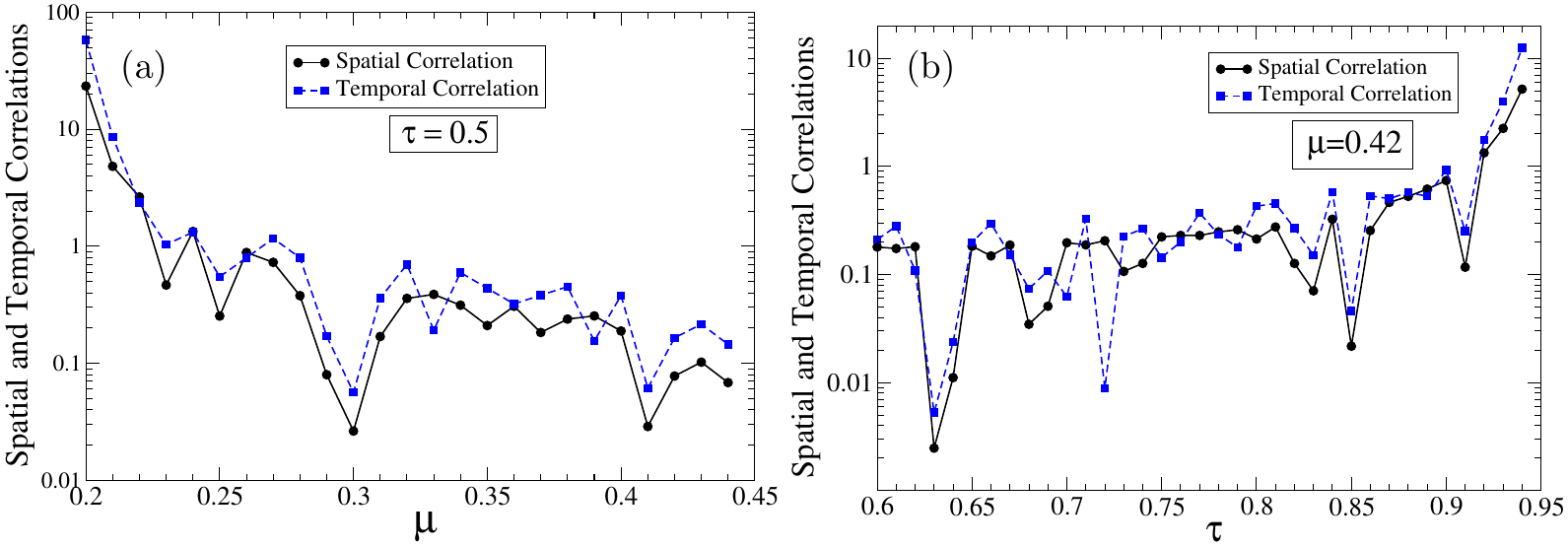}
\caption{Amplitudes of temporal and spatial correlation functions for
  $\mu=0.42$ as $\tau$ is varied (a) and for $\tau=0.5$ as $\mu$ is
  varied (b).  This figure complements the data shown in
  Fig.~\ref{figNEW333}. To quantify the oscillating amplitudes, we
  choose the average value of the amplitude in one period within the
  asymptotic regime for large $t'$ and $r$.  As $\mu$ is varied in
  (a), the amplitude of the correlation functions is low in the area
  of deterministically stable oscillations, but increases towards the
  limits of the stability region.  As $\tau$ is varied in (b), we
  observe qualitatively similar behavior: here the uniform
  oscillations lose stability as $\tau$ is lowered. The other
  parameters are as in Fig.~\ref{figNEW333}.}
 \label{figNEW555}%
\end{figure}

In order to obtain a more complete picture of the spatial and temporal
correlations, in Fig.~\ref{figNEW555}, we vary $\tau$ in small steps
for a fixed $\mu$ and vice versa. We have seen in Fig.~\ref{figNEW333}
that as we approach the stability boundary, the amplitude of the
correlation functions increases. For the temporal correlation
function, we average over the time interval $[(30-2\pi/\Omega),30]$,
and for the spatial correlation function, over the space interval
$[(50-2\pi/k),50]$, in order to ensure ergodicity over one full
period. As introduced above, $k$ denotes the most unstable wavenumber
of the uniform oscillations, and $\Omega$ their
frequency. Figure~\ref{figNEW555} shows that indeed the correlation
functions increase towards to the boundary where uniform oscillations
cease to be stable and standing waves set in the deterministic system.

\section{Spatio-temporal simulations in the presence of noise and feedback}
\label{sec:sim}

The expressions given in equations~(\ref{spatialcorr_eqn})
and~(\ref{tempcorr_eqn}) can be interpreted as a linear superposition
of two waves at the phase points $(k,x)$ and $(k,x+r)$ for all time
points, and with the same amplitude which is proportional to the noise
strength $D$. In other words, our model solution of the correlation
functions lead to noise-induced standing waves. In this section, we
show simulations that corroborate this. The amplitude of the Gaussian
noise term scales as $1/\sqrt{\Delta x \Delta t}$. This happens
because the two-point noise correlation is proportional to $\delta (x
-x')\delta (t -t')$. In the Euler discretization scheme, the additive
noise scales as $\sqrt{\Delta t}$.

\begin{figure}[h] 
 \centering 
\includegraphics[height=8.0cm,width=8.5cm]{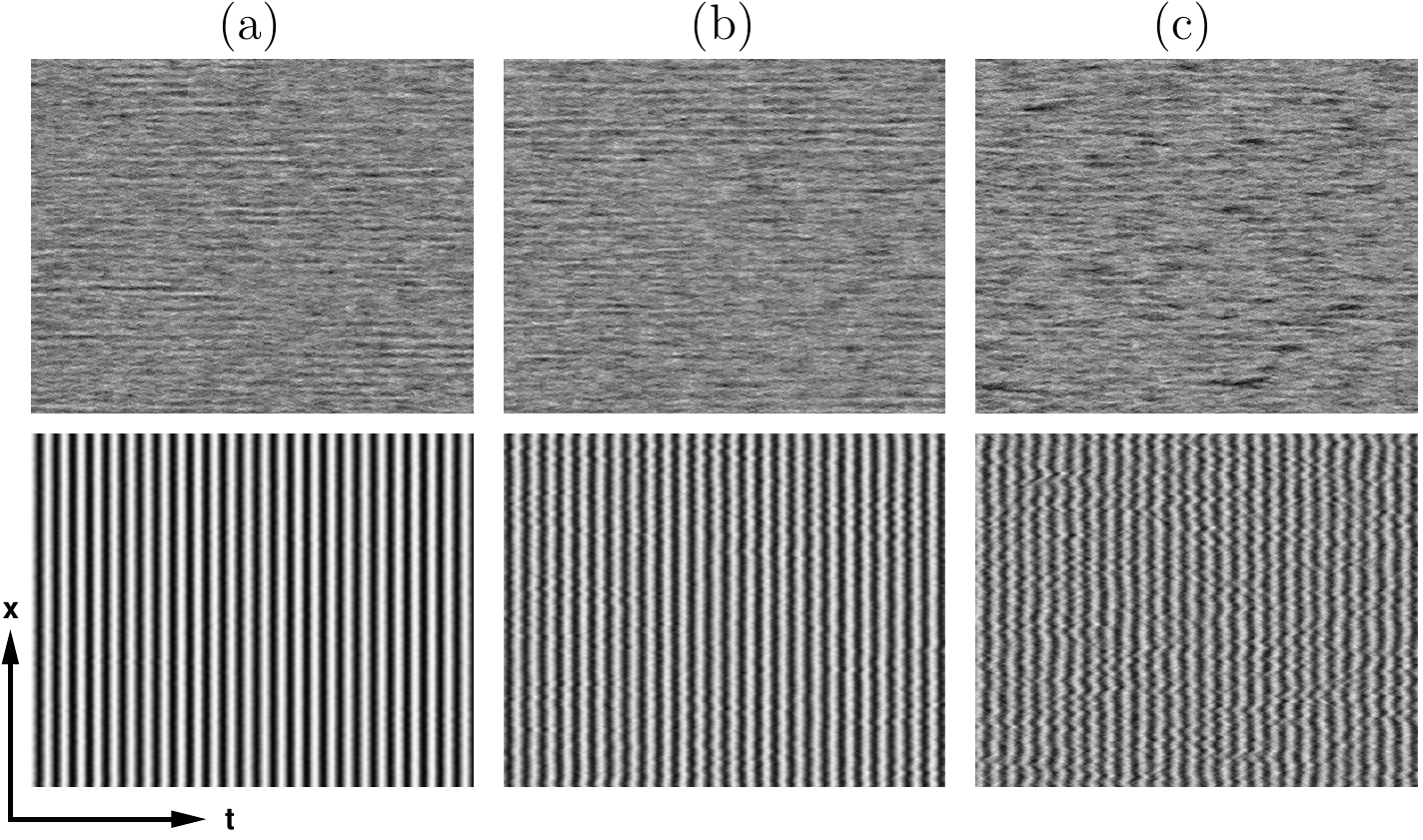}
\caption{Spatio-temporal simulations for different values of $D$ and
  fixed feedback strength. In space-time plots, we show $|A|$ (upper
  panels) and $ReA$ (lower panels) for $D=0.05$ (a), $D=0.2$ (b) and
  $D=0.5$ (c). The feedback magnitude is $\mu=0.25$, the delay time
  $\tau=0.5$, and the other parameters are as in
  Fig.~\ref{fig111}. For low $D$, we see that although $|A|$ shows
  some intermittent spatially periodic patches, their amplitude is
  actually quite small and the pattern actually is indistinguishable
  from uniform oscillations. For intermediate $D$, intermittent
  spatially periodic patches are seen in the pattern, reminiscent of
  standing waves. For large $D$, the noise is too strong to induce
  standing waves and the pattern corresponds to noisy uniform
  oscillations.}
 \label{fig555}%
\end{figure}

First, we consider a parameter value for which the deterministic
solution corresponds to uniform oscillations: delay time is fixed to
$\tau=0.5$ as above, and the feedback to $\mu=0.25$ which is larger
than the critical one, $\mu_c=0.19848$. In Fig.~\ref{fig555}, we show
three simulations, for increasing noise strengths. For $D=0.05$ (a) we
see an oscillatory pattern in the lower panel which is almost
indistinguishable from uniform oscillations. However, the upper panel
reveals that there is actually a spatial periodicity in $|A|$ and that
this periodicity is temporally persistent over multiple
oscillations. In the space-time plot, this is seen as patches of
horizontal stripes. This means that we observe a noise-induced spatial
pattern modulating the uniform oscillations, i.e., the formation of a
standing wave pattern. This finding resembles spatial
coherence~\cite{CarrilloEpL04}, as we will comment on below.

If the noise intensity is increased to $D=0.2$ (panel (b) of
Fig.~\ref{fig555}), we see similar patches of horizontal stripes in
the panel for $|A|$. However, their amplitudes are larger and
therefore, this time there is also a visible modulation of the
oscillatory pattern itself (lower panel of (b)). Hence, this pattern
corresponds to noise-induced standing waves. It is important to note
that the wavelength of the pattern corresponds to the
wavelength predicted through the linear stability analysis shown
in~\cite{BetaPD04,StichPD10}. This means, the wavenumber $k$
corresponds to the wavenumber $k_{max}$ for which $\lambda_1$ reaches
its maximum, while $\lambda_1(k_{max})<0$. If the noise intensity is
increased further to $D=0.5$ (c), patches of stripes give rise to more
irregular patches (upper panel). The lower panel shows oscillations
that are now visibly distorted by the noise, but without any spatial
periodicity.

\begin{figure}[h] 
 \centering 
\includegraphics[height=8.0cm,width=8.5cm]{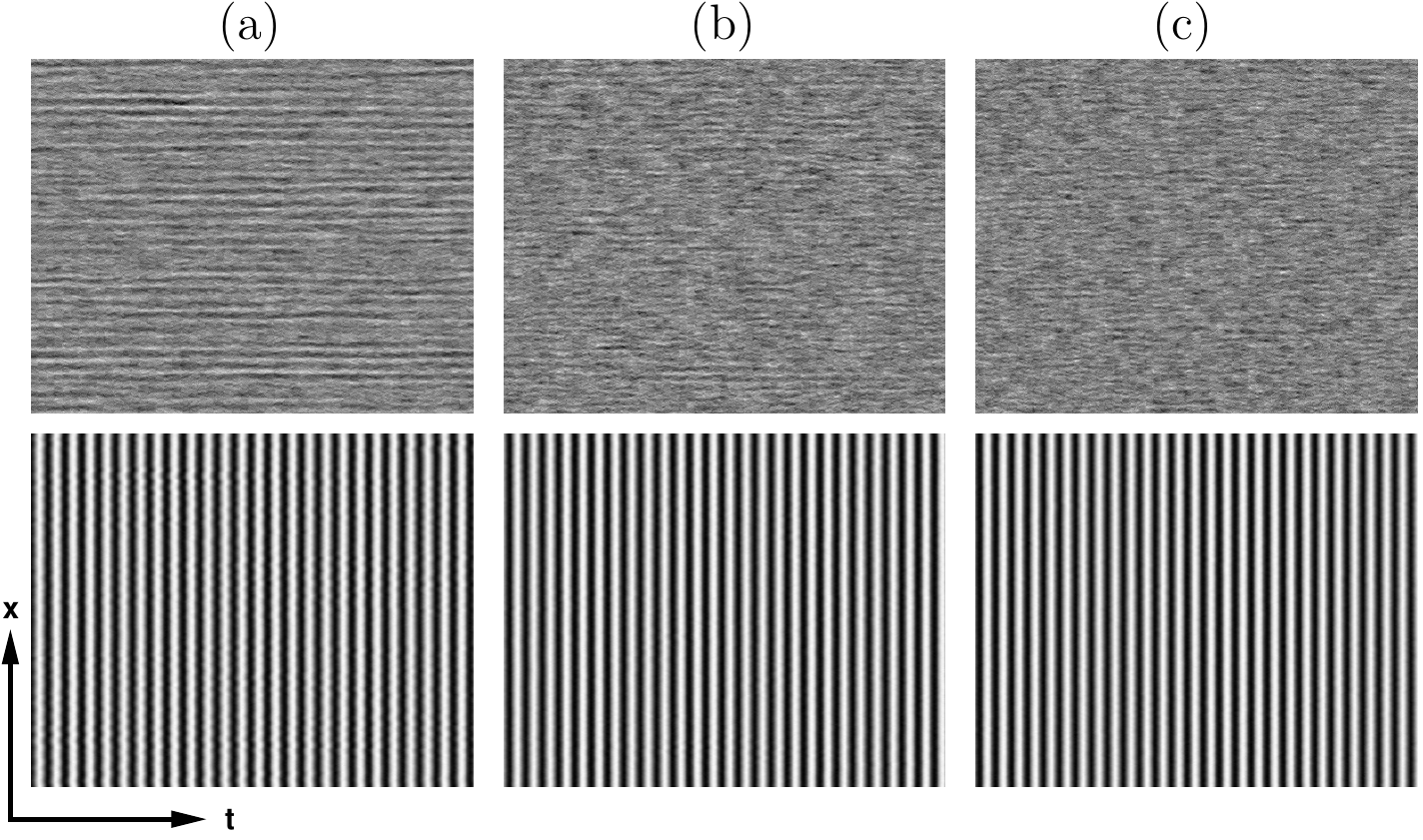}
\caption{Spatio-temporal simulations for different values of $\mu$ and
  fixed noise. In space-time plots, we show $|A|$ (upper panels) and
  $ReA$ (lower panels) for $\mu=0.2$ (a), $\mu=0.35$ (b) and $\mu=0.5$
  (c). The noise strength is $D=0.05$, the delay time $\tau=0.5$, and
  the other parameters are as in Fig.~\ref{fig111}. For low $\mu$, we
  observe clearly spatially periodic patterns that correspond to
  noise-induced standing waves. For larger $\mu$ (and hence further
  from the deterministic onset of standing waves), standing waves are
  weaker. For high $\mu$, the noise is not enough to induce standing
  waves.}
 \label{fig666}%
\end{figure}

We can now fix the noise intensity and explore the effect of varying
the feedback magnitude. In Fig.~\ref{fig666}, using $D=0.05$, we
display the results of spatio-temporal simulations for three values of
$\mu$ that all correspond to the regime where no standing waves are
stable in the deterministic system. First, we fix $\mu=0.2$ (a), a
value that ensures closeness to the onset of the standing wave
regime. Not surprisingly, we therefore see clear indication of
standing waves in the panel for $|A|$. However, similar to what has
been shown in Fig.~\ref{fig666}(a), the pattern amplitude is not large
enough compared to the uniform mode to be clearly seen in the
oscillations (lower panel).  Increasing the feedback magnitude to
$\mu=0.35$ (b), we see only weak evidence for patches of standing
waves (upper panel), and moving even further from the stability
boundary ($\mu=0.5$ in (c)), standing waves cannot be induced by weak
noise.

\begin{figure}[h]
\vspace{1cm} 
 \centering 
\includegraphics[height=5.0cm,width=8.5cm]{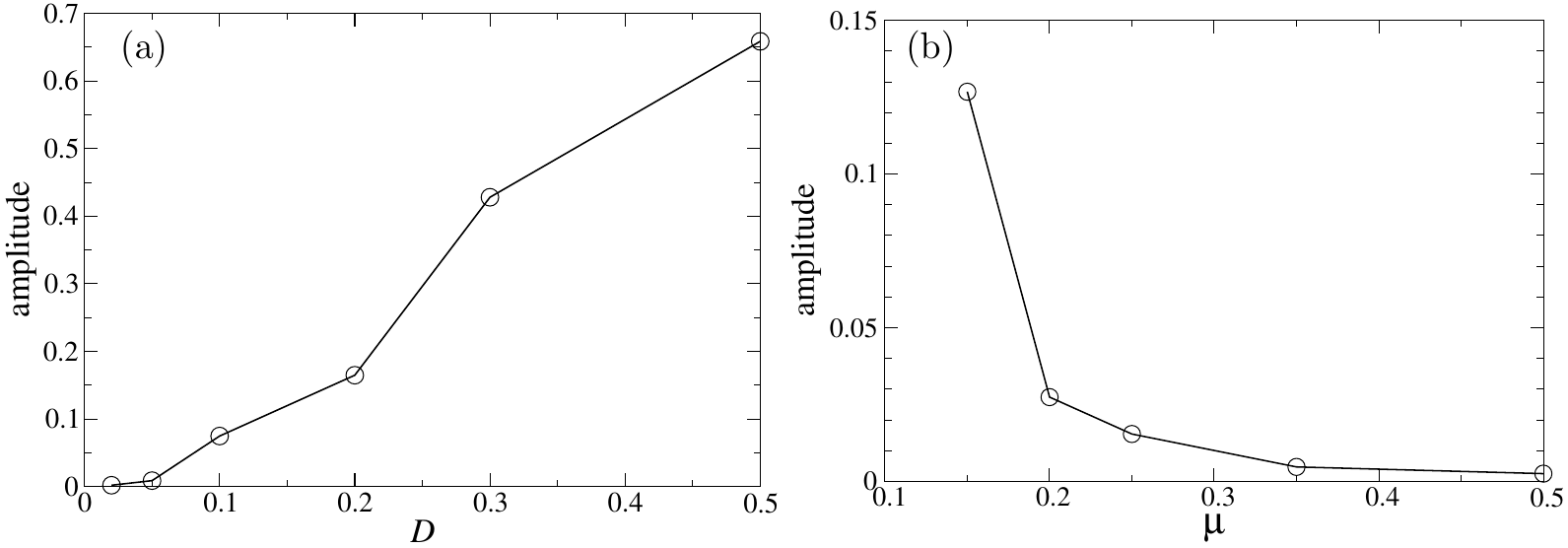}
\caption{Amplitude of noise-induced standing waves. In (a), this
  amplitude is shown as a function of the noise strength $D$ for a
  fixed set of $\tau=0.5$ and $\mu=0.25$. As qualitatively seen in
  Fig.~\ref{fig555}, for small $D$ the amplitude is very small and not
  perceivable, for intermediate $D$ the amplitude is large and
  visible, while for large noise strengths, the overall pattern
  becomes too irregular to actually identify standing waves. In panel
  (b), the amplitude of standing waves is shown as a function of the
  feedback strength $\mu$ for a fixed set of $\tau=0.5$ and noise
  strength $D=0.05$. The leftmost point ($\mu=0.15$) corresponds to
  deterministically stable standing waves with relatively large
  amplitude. Starting from $\mu=0.2$, we enter the regime where
  standing waves do not exist as deterministic solutions, and we see
  that the amplitude diminishes as we move away from the stability
  coundary of uniform oscillations.  The other parameters are as in
  Fig.~\ref{fig111}.}
 \label{figNEW888}%
\end{figure}

To assess the onset of noise-induced standing waves in more detail, we
obtain from the simulations (Figs.~\ref{fig555} and~\ref{fig666}) the
amplitude of the standing waves. To be precise, we show its spatial
contribution $2B_{k0}$ (see Eq.~(\ref{finalsol})), which should be
compared to the uniform contribution $H_0$, of order unity. Due to the
noisy character of the simulations, the standing waves occur only
intermittently and it is difficult to obtain their amplitude. In
Fig.~\ref{figNEW888}, we show how this amplitude varies with $D$ for
fixed $\tau$ and $\mu$ (a) and with $\mu$ for fixed $\tau$ and $D$
(b).

As seen in Fig.~\ref{figNEW888}(a), the amplitude increases
monotonically with $D$. The standing waves identified in the
simulations are in the intermediate parameter range: for small $D$,
uniform oscillations dominate, and for large $D$, the pattern becomes
very noisy on the background of uniform oscillations. In the scenario
shown in Fig.~\ref{figNEW888}(b), we observe a monotonically
decreasing amplitude profile with increasing $\mu$ that indicates
damping of the noise at large feedbacks.  As we deviate more and more
from the stability boundary, a given noise $D=0.05$ becomes more and
more ineffective to induce standing waves. Note that the first data
point ($\mu=0.15$) is already in the regime of deterministically
stable standing waves.

\section{Conclusion}
\label{sec:disc}

In this article, we studied standing waves for a complex
Ginzburg-Landau equation (CGLE) in the presence of global time-delay
feedback and noise and studied their properties analytically and
numerically. The CGLE describes the dynamics of a spatially-extended
system that undergoes a supercritical Hopf bifurcation. The basic
solution in this system corresponds to uniform oscillations. We
considered the situation where this solution is Benjamin-Feir unstable
in the absence of feedback ($1+\alpha\beta <0$), leading to
spatio-temporal chaos. Then, uniform oscillations or standing waves
can be induced through the time-delay feedback. Standing waves can be
understood as instability of the uniform oscillations, namely when the
oscillations become unstable with respect to perturbations with a
certain wavenumber (shown in Fig.~\ref{fig222}(b)). These waves
represent a transition state between uniform oscillations and a
chaotic state.

One main finding is that noise can induce standing waves in the regime
where uniform oscillations are stable (Figs.~\ref{fig555}
and~\ref{fig666}). The closer we are to the stability boundary that
separates uniform oscillations and standing waves, the less noise
intensity is needed to induce standing waves. If the system is at a
finite distance from that boundary, a comparatively larger magnitude
of the noise is needed to induce standing waves. In the limit $D\to
0$, no standing waves can be expected. However, as $D$ becomes large,
rather than inducing standing waves, irregular uniform oscillations
are observed. Hence, intermediate noise magnitudes are favorable for
the induction of standing waves. These results are similar in spirit
with findings of spatial or spatio-temporal coherence resonance
(e.g.,~\cite{CarrilloEpL04,PercPRE05,ZhouPRE02}). In contrast to those
works, however, we consider a system where the stable noise-free state
consists of uniform oscillations and the stabilized noise-induced
pattern consists of standing waves.  The wavenumber of the induced
standing waves agrees qualitatively with the value of $k$ for the most
unstable mode, as obtained by the stability analysis of uniform
oscillations. This is a common feature with pattern-forming systems
like the one discussed in~\cite{CarrilloEpL04} due to the appearance
of an intrinsic length scale.

For the noisy CGLE and in absence of feedback, standing waves have not
been reported. So feedback is still essential for finding standing
waves. However, we emphasize that the onset of standing waves can be
controlled by noise. The CGLE represents an oscillatory
reaction-diffusion system where the chaos is diffusion-induced and
hence there is a fundamental difference to the oscillators
in~\cite{ZhouPRE02,KissC03} which display a chaotic dynamics without
coupling and where phase synchronization of oscillations (and no
standing waves) are observed.

The correlation functions evaluated in the regime of deterministically
stable uniform oscillations (Fig.~\ref{figNEW333}) show oscillations
that increase while approaching the deterministic stability boundary,
corroborating the idea of noise-induced standing waves in this
parameter regime. More generally, we note that noise does not destroy
the deterministic Hopf bifurcation structure itself but only modulates
the instability leading to standing waves.  We have verified this for
the range of parameter values studied, i.e., for small delays $\tau\le
1$ and moderate feedback magnitudes $\mu\le 1$. Future work will
target different (wider) regimes.

We showed that small noise does not destabilize deterministically
stable standing waves (Fig.~\ref{fig222}(c)), but we have not studied
systematically what effect noise exerts on standing waves where these
are stable in the deterministic system and on the chaotic solution
itself.  Future work may comprise a study to characterize these
dynamics and separate it from spatio-temporal chaos that is found when
the feedback strength is decreased in the deterministic system.


\section{Acknowledgments}

MS acknowledges stimulating discussions with Carsten Beta and Eckehard Sch\"oll.



\end{document}